\documentclass[rmp,twocolumn,preprintnumbers,amsmath,amssymb]{revtex4}
\usepackage{dcolumn}% Align table columns on decimal point
\usepackage{bm}% bold math

\newcommand{\beq}{\begin{equation}}
\newcommand{\eeq}{\end{equation}}
\newcommand{\bea}{\begin{eqnarray}}
\newcommand{\eea}{\end{eqnarray}}
\newcommand{\uvec}[1]{{\bf \hat{#1}}}
\newcommand{\eq}[1]{{Eq. (\ref{#1})}}

\usepackage{amssymb}
\usepackage{graphicx}
\begin{document}
\preprint{APS/123-QED}

\title{Bifurcation and chaos in spin-valve pillars in a periodic applied
magnetic field }
\author{S. Murugesh{$^1$}}
\email{murugesh@phy.iitkgp.ernet.in}

\author{M. Lakshmanan{$^2$}} 
\email{lakshman@cnld.bdu.ac.in}
\affiliation{$^1$Department of Physics \& Meteorology, IIT-Kharagpur,
  Kharagpur 721~302, India}
\affiliation{$^2$Centre for Nonlinear Dynamics, School of Physics, 
  Bharathidasan University, Tiruchirapalli 620 024, India}

\date{\today}

\begin{abstract}
We study the bifurcation and chaos scenario of the macro-magnetization vector 
in a homogeneous
nanoscale-ferromagnetic thin film of the type used in spin-valve pillars.
The underlying dynamics is described by a generalized Landau-Lifshitz-Gilbert
(LLG) equation. 
The LLG equation has an especially appealing form under a complex
stereographic projection, wherein the qualitative equivalence of
an applied field and a spin-current induced torque is transparent. Recently 
chaotic behavior of such a spin vector has been identified by Zhang and Li 
using a spin polarized current passing through the pillar of constant 
polarization 
direction and periodically varying magnitude, owing to the spin-transfer
torque effect. In this paper we show that the same dynamical
behavior can be 
achieved using a periodically varying applied magnetic field, in the presence
of a constant DC magnetic field and constant spin current, which is 
technically much more feasible, and demonstrate numerically the chaotic
dynamics in the system for an infinitely thin film.
Further, it is noted that in the presence of a nonzero crystal anisotropy
field chaotic dynamics occurs at much lower magnitudes of the spin-current
and DC applied field.

\end{abstract}
%\pacs {75.10.Hk, 67.57.Lm, 75.60.Jk, 72.25.Ba }
\maketitle
{\bfseries \noindent
Bolstered by the importance of Giant Magneto Resistance (GMR),
a sequence of experimental and theoretical developments in the
last few years on current induced switching of magnetization in nanoscale 
ferromagnets has thrown open several prospects in next generation magnetic 
memory devices. The direct role of spin polarized current, as against the
traditional applied field, in controlling spin dynamics has brought in 
the possibility of new types of current-controlled memory devices and 
microwave resonators. The system under consideration is primarily a nanoscale
spin-valve pillar structure, with one {\it free} ferromagnetic layer and 
another {\it pinned} layer separated by a non-ferromagnetic conducting layer. 
The behavior of the dynamical quantity of interest, the magnetization field
in the free layer, is well modeled by an extended Landau-Lifshitz 
equation with Gilbert
damping, which is a fascinating nonlinear dynamical system. The free layer
is usually assumed to be of single magnetic domain. Owing to the highly 
nonlinear nature of the LLG equation it is imperative to study the chaotic
dynamical regime of the magnetization field. Indeed, several recent 
experiments have exclusively focused on chaos aspect of the system. In this
article we have shown that a small applied periodically varying (AC) magnetic 
field, in the presence 
of a constant spin-current and a steady applied magnetic field, can induce 
parametric regimes 
displaying a broad variety of dynamics and period doubling route to chaos. 
A numerical study of the effects of a non-zero anisotropy field reveals 
chaotic dynamics at much lower magnitudes of the spin-current and applied
DC field. This could be an important factor to consider in 
microwave resonator applications of spin-valve pillars.  }

\section{Introduction} 
Following the success of GMR, the next major development in classical computer 
technology is expected to be through MRAMs
\cite{wolf:2006,nesbet:1998,tsang:1998}. Apart from a manifold reduction in  
power consumption, being inherently nonvolatile in nature, they also bring 
in the prospect of computers that need not be rebooted. Understandably, 
fabrication
of MRAMs has been a major thrust area of research in the last two decades. 
Typically, the memory unit consists of two nanoscale magnetic films separated
by a spacer conductor/semiconductor medium and works on the principle of GMR.
%When a current is passed between the two magnetic films, the resistance 
%depends on the relative magnetization between the two films. Usually the
%resistance is less when the magnetization vectors are parallel. The direction
%of magnetization of one of the films is kept fixed, while that of the
%other is flipped, usually by an applied magnetic field. This effectively
%is the process of {\it writing}, while the application of the current,
%and the subsequent measurement of the resistance serves the purpose of 
%{\it reading} data. 
The imminent
prospects in the recording media industry has prompted a breadth of 
development in the field. Besides the eminent role as a memory unit, the 
possibility of a single MRAM as a logical unit has also been proposed 
\cite{black:2000,koch:2003}.  
One notable technological hiccup in fabricating large MRAM grids is the
extent to which the applied magnetic field can be localized. This imposes
limitations on the efficiency with which an individual unit can be
manipulated. The applied field required for the purpose is in most cases 
the Oersted field generated through electrical currents. A significant step
forward, in bypassing the limitation on field localization, occurred when
Slonczewski and Berger independently envisaged a more direct role for
(spin polarized) current on magnetization\cite{slonc:1996,berg:1996}. 
They predicted that the 
angular momentum acquired by the spin polarized current can interact
with the magnetization vector, and thus a suitable spin polarized current can 
possibly flip its direction. This phenomenon has been well established in a 
series of experiments in the last decade and referred to as the 
{\it spin-torque effect} in the literature\cite{stiles:2006,berk:2008,myers:1999,wegrowe:2001,kat:2000,gro:2001,kent:2003,bass:2003}. 
Interestingly, the effect has a simple semiclassical description in the  
form of an extended Landau-Lifshitz equation with Gilbert damping
\cite{berg:1996,slonc:1996,baz:1998}. 
Several proposals have appeared in the
last few years suggesting an increased role of the spin current and the
associated torque in the eventually expected version of the MRAM.   

One important assumption often made in most of these studies is that the 
magnetization in the film is homogeneous, or at least enough well defined that
one can consider the film to be a mono-domain layer. This homogeneity 
assumption effectively nulls
the Heisenberg interaction between neighboring spins, and allows one to treat 
the system as a single spin unit. As the size of the magnetic film is 
increased,
this approximation is expected to fail. 
%Indeed, chaotic behavior has been
%observed at lateral sizes of order $60-130~ nm$\cite{kise:2003,lee:2004}. 
%A rapid quenching, or flipping,
%of the magnetization is another scenario where the interaction between 
%neighboring spins
%assumes more significance, and can break the inhomogeneity. 
%Besides, it is well realized experimentally that spin-transfer can induce
%microwave oscillations\cite{weber:2001,kise:2003,Xi:2004,ripp:2004}. The 
%possible role of such current controlled 
%microwave oscillators in the nanoscale has been well realized
%\cite{kise:2003}. For higher power levels that are practically desired
%it is natural to look at a series of such coupled spin-valve oscillators. 
%Studies on synchronization of such coupled oscillators in the spin-valve
%geometry induced by spin-transfer torque, and modeled on the extended
%Landau-Lifshitz equation have been carried out both experimentally and
%numerically \cite{pers:2007,fert:2006,kaka:2005,manc:2005}.  
%The prospect of an AC generator in the nanoscale by
%application of a steady spin-current apart, chaotic behavior itself
%may have direct application. Synchronization of chaotic
%behavior in coupled oscillators has immense applications in communication
%and cryptography that are well reported \cite{mlsr:2003}.
%Under these circumstances,
%further study of chaotic dynamics in spin-valve structures becomes inevitable. 
Indeed, chaotic behavior has been observed due to spin field inhomogeneity
at lateral sizes of order $60-130~nm$ \cite{kise:2003,lee:2004}. 
Besides, it is well realized experimentally that spin-transfer can induce
microwave oscillations\cite{weber:2001,kise:2003,Xi:2004,ripp:2004}. The 
possible role of such current controlled 
microwave oscillators in the nanoscale has been well realized
\cite{kise:2003}. For higher power levels that are practically desired
it is natural to look at a series of such coupled spin-valve oscillators. 
Studies on synchronization of networks of such coupled nano-spin transfer 
oscillators, each one modeled on the extended LLG equation
have been carried out both experimentally and
numerically in an effort to enhance the emitted power 
\cite{manc:2005,kaka:2005,fert:2006,pers:2007}.  

Since the extended LLG equation is highly nonlinear in nature, a detailed
study of the underlying 
nonlinear dynamics in spin-valve structures becomes inevitable.
The stability diagram based on Melnikov theory in the space of the external 
field along the direction of anisotropy and the strength of spin torque 
due to a DC current was obtained in detail by Bertotti {\it et. al.,} in
\cite{bert:2005}. Following this, it was shown by Z. Li, Y. C. Li and S. Zhang
that an applied 
alternating  current can induce three broad types of dynamics, $vis-a-vis$
Chaos, Multiply periodic and Periodic \cite{charles:2006}. Qualitatively 
similar features are also noted when the effects of nearest neighbor 
interactions are included in a long one dimensional ferromagnet with
a spin torque term. Indeed the ferromagnetic chain exhibits both
periodic and chaotic behavior in the presence of an applied AC spin current.
\cite{lan:2008}.  The multiply
periodic behavior refers to the case where, with change in parameter,
the system moves from periodic to multiply periodic behavior but not
leading to chaos. It must be
mentioned at this point that the last two types `multiply periodic' and
`periodic' are also referred to in the literature as `modification' and 
`synchronization'\cite{charles:2006,yang:2007}. The usage here is deliberate 
as the periods do not have a direct relation to the period of the applied
magnetic field. It was also
shown that in the space of the applied external field and the strength
of the spin current the system exhibits chaotic behavior for a range
of values within the boundary predicted by Melnikov theory \cite{yang:2007}.

The possibility of chaotic oscillations in monodomain single nanoscale 
spin-valve requires further in depth study of underlying bifurcation and 
chaos scenarios, including the detailed phase diagrams, at least for two 
reasons: (i) Chaotic oscillations in spin transfer oscillators may be 
unfavorable from a practical/technological point of view and such 
oscillations need to be suppressed by minimal intervention using one of 
the several controlling techniques developed in the field of chaotic dynamics 
\cite{mlm:1995,mlsr:2003}, where already 
such a study has been made \cite{xu:2008}. (ii)  
The previously mentioned possibility of synchronization of coupled spin-valve 
oscillators raises complex problems which are new in spintronics and is
related to the topic of chaos synchronization in dynamical systems 
\cite{mlm:1995,mlsr:2003,piko:2008}, similar to 
synchronization of chaotically evolving networks of Josephson junctions, 
laser systems and nonlinear oscillators.  As synchronization of chaotic 
oscillators is considered as a potential technique for secure communication 
including cryptography,  there exists possible applications of networks 
of nanoscale spin-valve structures along these lines, although this may be
complicated due to the presence of several parameters in the system.  
Consequently the study 
of the full nonlinear dynamics of spin transfer oscillators will be of 
considerable significance.
It may be noted that chaos in magnetic systems, such as yttrium garnet, 
driven by external fields have been extensively studied in the past 
\cite{rez:1990,carr:1991,pisk:1999}. Numerical experiments on a model
of thin magnetic film wherein spin wave excitations induced by spin-current
lead to chaos have also been studied in \cite{aug:2002}. 
However, chaos in nano-spin valve geometry is reasonably new with potential 
new applications as discussed above.

In this paper, we study the chaotic dynamics of the magnetization vector
in a single domain current driven spin-valve pillar, induced by a periodically
varying (AC) applied magnetic field in the presence of a constant spin current
and steady (DC) magnetic field, using the extended 
LLG equation
as the model for the system. Making use of a complex stereographic variable, 
we observe that the spin current induced torque is effectively equivalent to 
an 
applied magnetic field. Following this observation, we show numerically that
a periodically alternating field can lead to chaotic behavior of the
magnetization vector, which is similar to that of an alternating spin-current
induced torque, studied in Li, Li, and Zhang \cite{charles:2006}. It will be 
shown that the order of magnitude of 
the applied alternating field required for chaotic motion is substantially 
lower, within practically achievable limits, compared to the 
alternating current magnitudes shown in 
\cite{charles:2006,yang:2007}. This is expected to be helpful in applications
such as resonators, as an AC magnetic field is much more feasible practically
than a AC spin-polarized current (although, when it comes to DC fields,
the current induced DC Oersted fields are more cumbersome to produce in 
spin valve geometries compared to DC spin currents). 
We study the dynamics in an infinite thin film, both with a vanishing and
non-vanishing anisotropy field. In the setup we shall consider the
monodomain spin layer influenced by a DC spin-current, and both a DC and
AC applied magnetic field. Although the applied magnetic field is qualitatively
equivalent to a suitable spin-current\cite{sm:2008}, we employ both a 
DC spin-current and applied magnetic field as the magnitudes at which chaos is
observed is quite high, and hence difficult to achieve exclusively using
either of the two. 
It may be noted that the chaotic dynamics studied here is induced
by an AC applied magnetic field, and is phenomenologically different 
from the spin field
inhomogeneity induced chaotic behavior that is observed
in \cite{kise:2003}. Periodic dynamics is possible even in the absence
of an alternating current, or field, as has been noticed in ferromagnetic
films induced by a spin-current \cite{ripp:2004}. Futher it has been reported
therein that the current magnitudes at which periodic behavior is seen 
share a linear relation of negative slope with the oscillation frequencies. 
Our numerical
results based on the extended LLG model are shown to be in agrement with these
observations. Interestingly, in the presence of a non-zero
in plane easy-axis crystal anisotropy field (taken along the $z$ direction),
the chaotic dynamical regime is observed at much lower magnitudes of the DC
applied field and spin-current. This could be an aspect to factor in 
while designing spin-valve based microwave resonators. 
%It is further
%noted that for certain low values of the applied DC field and current
%there appear large regions of periodic behavior that provide a better and
%stable choice in microwave resonator applications. Chaos in 
%magnetic systems, driven by external fields and currents have been 
%extensively studied in the past \cite{rez:1990,carr:1991,pisk:1999}. However,
%chaos in nano-spin-valve geometry is reasonably new with immense potential. 
%With the application
%potential in technological developments in the near future being immense,
%further study on the 

The paper is organized as follows. In Section 2, we detail the various
interactions, including spin-transfer torque, that make up the extended LLG
equation. Further, we introduce a complex stereographic variable
equivalent to the macro-magnetization vector, and rewrite the LLG
equation in this variable. As will be shown, this elucidates the role of the
spin transfer torque as equivalent to an applied magnetic field. In Section 3,
we present our numerical results which demonstrate chaotic dynamics of the
magnetization vector in the presence of a periodic applied field. Here we
shall consider two cases, namely response in the presence and absence
of a crystal anisotropy field. As will be noticed, the chaotic regimes
in the two cases are significantly different. In Section 4
we conclude with a discussion on our results. 

\section{The extended Landau-Lifshitz equation and complex representation}\label{2}

\begin{figure}[h]\centering\includegraphics[width=1\linewidth]{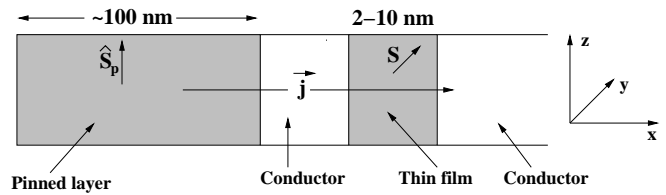}
\caption{A schematic figure of a spin-valve pillar. The cross section of the
free layer is roughly $5000~nm^2$. $\uvec{S}$ is the magnetization vector
in the free layer, and is the dynamical quantity of interest. $\uvec{S}_P$
is the direction of polarization of the spin current.}
\end{figure}

A typical spin-valve pillar used in most experiments is schematically shown
in FIG 1. 
A current passing 
through this ferromagnet acquires a spin polarization in the $\uvec{z}$
direction. The thickness of the spacer conductor medium should be
less than the spin diffusion length of the polarized current. The 
polarized current then passing through the free layer causes a change 
in the magnetization vector $\uvec{S}$, an effective torque referred to as the
spin transfer torque.  
Interestingly, it has been realized that semiclassically the 
phenomenon can be described by an extension to the LLG equation,
\cite{slonc:1996,berg:1996}
\beq\label{ll}
\frac{d{\uvec{S}}}{dt} = -\gamma{\uvec{S}}\times \vec{H}_{eff} 
+ \lambda\uvec{S}\times\frac{d\uvec{S}}{dt}
- \gamma~a\uvec{S}\times(\uvec{S}\times\uvec{S}_p),
\eeq
Here, $\uvec{S}=\{S_1,S_2,S_3\}$ is the unit vector along the direction of the 
magnetization vector in the ferromagnetic film, which is the dynamical
variable of interest, $\gamma$ the gyromagnetic ratio,  and $\lambda$ the 
dissipation coefficient. $\vec{H}_{eff}$ in \eq{ll} is the effective field
due to exchange interaction, anisotropy, demagnetization and an applied field
(see \cite{sm:2008} for details):
\bea\label{heff}
\vec{H}_{eff} = DS_0\nabla^2\uvec{S} +
\kappa(\uvec{S}\cdot\uvec{e}_{\parallel})\uvec{e}_{\parallel}
+\vec{H}_{demag}
+ \vec{B}_{applied},
\eea
$\uvec{e}_{\parallel}$ being the unit vector along the direction of the
anisotropy axis. 
The demagnetization field is obtained as a solution of the differential
equation
\beq\label{demag}
\nabla\cdot\vec{H}_{demag} = -4\pi S_0\nabla\cdot\uvec{S},
\eeq
and $\vec{B}_{applied}$ is the applied magnetic field on the sample. 
The last term in \eq{ll} is the additional term describing the spin-transfer
torque, and the parameter `$a$' depends linearly on the current density $j$. 
$\uvec{S}_p$ 
is the direction of magnetization of the polarizer, i.e., the polarization 
of the spin current.

In this study we shall
assume the magnetization to be homogeneous. This effectively nulls the 
exchange interaction, while \eq{demag} can be immediately 
solved to give the demagnetization field as 
\beq
\vec{H}_{demagnetization} = -4\pi S_0( N_1S_1\uvec{x} + N_2S_2\uvec{y} 
+ N_3S_3\uvec{z}),
\eeq
where $N_i,i=1,2,3$ are conveniently chosen such that  $N_1+N_2+N_3=1$ 
(after suitable rescaling of the magnitude of the spin). 
For a spherical sample $N_1=N_2=N_3$, and 
the demagnetization term is effectively null in \eq{ll}. 
In the next sections we study
chaotic dynamics shown by the system when an alternating magnetic field is 
applied. 

%\section{Complex representation using stereographic variable}\label{3}
Rewriting \eq{ll} using the complex stereographic variable 
\cite{naka:1984,kosaka:2005} 
\beq\label{stereo}
\Omega \equiv \frac{S_1+iS_2}{1+S_3},
\eeq
provides a more comprehensible picture of the role of spin transfer torque. 
For the spin valve system, the direction 
of polarization of the spin-polarized current $\uvec{S}_p$ remains
a constant, and lies in the plane of the film. 
Without loss of generality, we call this the direction
$\uvec{z}$ in the internal spin space, i.e., $\uvec{S}_p = \uvec{z}$. 
As mentioned in Sec. 2, we disregard the exchange term. We choose the applied
external field also in the $\uvec{z}$ direction, i.e., 
$\vec{B}_{applied}= \{0,0,h_{a3}\}$. Defining
\beq\label{ep}
\uvec{e}_{\parallel} = \{\sin\theta_\parallel\cos\phi_\parallel,
\sin\theta_\parallel\sin\phi_\parallel,\cos\theta_\parallel\}
\eeq
and upon using \eq{stereo} in \eq{ll}, we get 
\begin{widetext}
\bea\label{llo}
(1-i\lambda)\dot{\Omega} = 
-\gamma(a-i h_{a3})\Omega 
+ iS_{\parallel}\kappa\gamma\Big{[}\cos\theta_\parallel\Omega 
-\frac{1}{2}\sin\theta_\parallel(e^{i\phi_\parallel} - %\nonumber\\
\Omega^2e^{-i\phi_\parallel})\Big{]}
-\frac{i\gamma4\pi~S_0}{(1+|\Omega|^2)}\Big{[}N_3(1-|\Omega|^2)\Omega\nonumber\\
-\frac{N_1}{2}(1-\Omega^2-|\Omega|^2)\Omega%\nonumber\\
-\frac{N_2}{2}(1+\Omega^2-|\Omega|^2)\Omega 
- \frac{(N_1-N_2)}{2}\bar{\Omega}\Big{]},
\eea
\end{widetext}
where $S_{\parallel} = \uvec{S}\cdot\uvec{e}_{\parallel}$. Using \eq{stereo} and 
\eq{ep}, $S_{\parallel}$, and thus \eq{llo}, can be written entirely in terms 
of $\Omega$. For further details of derivation of \eq{llo} see Ref. 
\cite{sm:2008}. 

It is interesting to note that in this representation the spin-transfer torque,
proportional to the parameter $a$, appears only in
the first term in the right hand side of \eq{llo} as an addition to the
applied magnetic field $h_{a3}$ but with a prefactor $i$. Thus the spin
torque term can be considered as an effective applied magnetic field 
\cite{sm:2008}. It was further explicitly shown in \cite{sm:2008}, in
the absence of the crystal anisotropy field, that the switching time due to
the spin-torque will be shorter by a factor $\lambda$, compared to that of 
magnetic field induced switching. Further, the spin-torque produced the
dual effect of precession and dissipation even in the presence of the external
applied field. 
The nature of switching of magnetization for other types of materials can be
investigated by analyzing \eq{llo}, and for typical materials this has been
carried out in \cite{sm:2008}.

\begin{figure}[h]\centering\includegraphics[width=.8\linewidth]{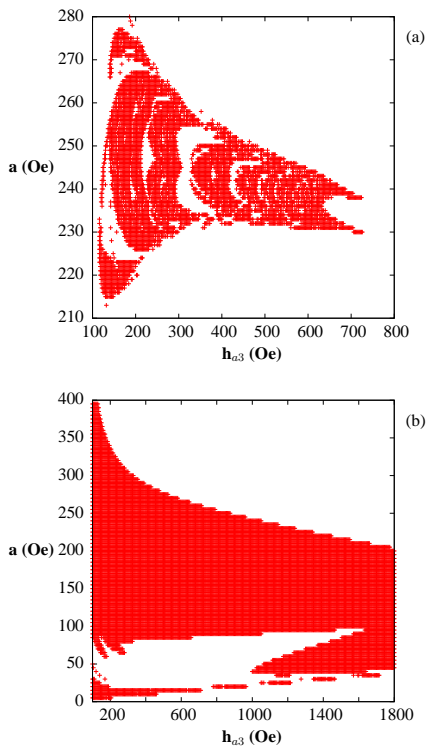}
\caption{Regions of chaos in the $a-h_{a3}$ space, for an applied 
alternating magnetic field of amplitude $h_{ac}=10~Oe$ and frequency 
$\omega=15~ns^{-1}$, a) without anisotropy field, $\kappa=0$, and b) with 
anisotropy field of strength
$\kappa=45~Oe$ along the $z$ direction. 
The dark regions indicate values for which the dynamics is chaotic, i.e, 
regions where the largest Lyapunov exponent is positive. In a) chaos is 
rarely noticed for lower values of $h_{a3}$. The other parameters
are $N_1=1$, $N_2=N_3=0$, $4\pi S_0=8400 Oe$. The points are plotted at
intervals of $5~Oe$ along both axes, and hence the figure offers only limited
resolution in the dark(chaotic) regions.  }
\end{figure}

\section{Chaotic dynamics}
Magnetization reversal in a spin mono-domain layer in the presence of both
a steady applied magnetic field and a steady polarized current corresponds 
to a rather 
complicated phase diagram, as revealed in \cite{bert:2005}. Using Melnikov
theory, it was also shown therein that the magnetization vector also has
limit cycles for a range of values of the parameters, with frequency in 
the microwave range. 
The dynamical quantity in question, the magnetization  vector
$\uvec{S}$, is two dimensional, owing to its constant (unit) magnitude. 
Hence, chaotic behavior is ruled out. However,  making the applied field,
or current, time dependent is one way of increasing the effective 
dimensionality of the system to three and hence introduce a possibility of 
chaotic dynamics. Following \cite{bert:2005} it has been shown in 
\cite{charles:2006} that a small alternating current can produce a variety of 
dynamics, namely, {\it Multiply periodic, Periodic} and {\it Chaos}. It is
also noticed that, along with a steady spin current of order
$250~Oe$ and a steady applied magnetic field of the same order, 
inclusion of a small {\it alternating} spin polarized current leads to 
chaotic dynamics \cite{yang:2007}. 

The dynamical similarity 
of the applied field and the spin-torque was noted in Section \ref{2}. 
In this section we show numerically that an applied AC field can also produce 
diverse dynamical scenarios and point out the advantages in using a 
periodically varying 
applied magnetic field instead of an alternating current. i.e., in \eq{ll} (or
equivalently \eq{llo}) we take 
\beq
\vec{B}_{applied} = \{0,0,h_{a3}+h_{ac}\cos\omega t\}.
\eeq
It will be noted that chaotic dynamics is possible
at much lower DC applied field strengths and spin current in the presence
of an anisotropy field. 

For the film geometry, the film is taken to be in the $y-z$ plane, and the 
demagnetization field perpendicular to the film, the $\uvec{x}$ direction, i.e.,
\beq
H_{demagnetization} = -4\pi S_0S_1\uvec{x}.
\eeq
The saturation magnetization is taken to be that of permalloy, so that
$4\pi S_0=8400 Oe$. Two different scenarios, one without anisotropy, and
another with an in-plane easy axis anisotropy of strength $\kappa=45~Oe$ 
along the $z$ direction, are investigated.  All the numerical results that 
follow have been obtained by directly simulating \eq{ll} for the vector
$\uvec{S}$. The same results were confirmed using \eq{llo} as well.

\begin{figure}[b]
\centering\includegraphics[width=1\linewidth]{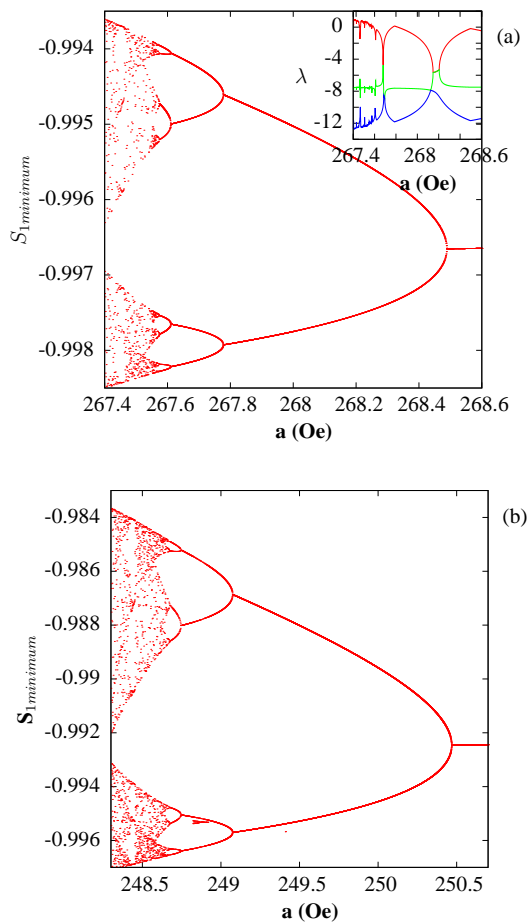}
\caption{Period doubling route to chaos as $a$ is varied. The figure is a plot
of the minimum values of $S_1$ over several periods for the given parameter
values a) without anisotropy, and b) with anisotropy of $\kappa=45~Oe$ along
the $z$ direction. The applied DC field 
$h_{a3}=200~Oe$. All the other parameters remain the 
same as in FIG 2. The corresponding Lyapunov spectrum is shown as an inset.}
\end{figure}

\subsection{Regions of Chaos in the presence of an applied alternating field}
As a first step we show below (FIG 2) the regions of chaos, or regions of
positive Lyapunov exponent, in the space of the DC magnetic field and the DC 
current for 
an alternating applied magnetic field of magnitude $10~Oe$ and frequency 
$15~ns^{-1}$, 
first without an anisotropy field, $\kappa=0$, (FIG 2a), and then with 
anisotropy field, $\kappa=45~Oe$,
(FIG 2b). The regions are obtained by direct numerical simulation using
the model in \eq{ll}, or equivalently \eq{llo}. The dark regions in FIG 
2 represent parameter values when the largest Lyapunov exponent is positive. 

%here.

\subsubsection{Case a: No anisotropy ($\kappa=0$)}
The similarity of FIG 2a with that of regions of chaos for an alternating 
spin current must 
be noted (see FIG 1 in \cite{yang:2007}). The figure is a demonstration
of the qualitative equivalence of the applied field and current induced
spin-torque in describing the gross dynamical scenario. From FIG 2a,
chaotic behavior of spins is observed for spin-current magnitudes in the 
range of 
$a=200-300~Oe$, and DC magnetic fields above $100~Oe$. These values of `$a$'
correspond to spin-current magnitudes of over $10^{12}A/cm^2$, which is at the 
higher end of the presently achievable levels. 

In FIG 3 we present the bifurcation diagram showing the period doubling 
route to chaos as the DC current is varied. These diagrams are obtained 
by plotting the 
minimum values for one of the components of the spin, in this case $S_1$, 
over several periods of time for each value of $a$ in the given range. 
From the data in FIG 3a, the first five period doubling bifurcations are 
seen 
to occur at $a_n=268.4845,267.7723,267.6055,267.5685,267.5605$. 
Consequently, the ratio $\delta_n = (a_n-a_{n-1})/(a_{n-1}-a_{n+1})$ takes 
values $4.2698,4.5081,4.625$, clearly approaching the Feigenbaum constant.
The Lyapunov spectrum for this range of $`a'$ is shown as inset.
On comparison, we notice the largest Lyapunov exponent is positive for 
values of $a$ where the dynamics is chaotic in FIG
3a. A similar check has been made for FIG 3b, which again shows the period
doubling route in the presence of an anisotropy field. 
%The power spectrum corresponding to periodic and chaotic dynamics 
%in FIG 3a are shown in FIG 4. 
Period doubling route to chaos is also noticed as the strength 
of the steady applied magnetic field is varied(keeping the current and 
frequency fixed),  and when the frequency of the applied field is varied
(keeping current and magnetic field strength steady), respectively.

%\begin{figure}[h]
%\centering\includegraphics[width=0.8\linewidth]{fig6.eps}
%\caption{Periodic doubling route to chaos as seen in the presence of
%a nonzero anisotropy field of strength $\kappa=45~Oe$ along the $z$ direction.
%(a) the DC magnetic field $h_{a3}$ is varied, keeping the frequency constant
%at $\omega=15~~ns^{-1}$, and (b) with change in frequency 
%$\omega$ of the applied AC
%magnetic field, keeping the DC magnetic field constant at $20~Oe$. In 
%either case the magnitude of the DC spin current is maintained constant
%at $150~Oe$. All the other parameters remain the same as in FIG 2. }
%\end{figure}

\begin{figure}[h]
\centering\includegraphics[width=1\linewidth]{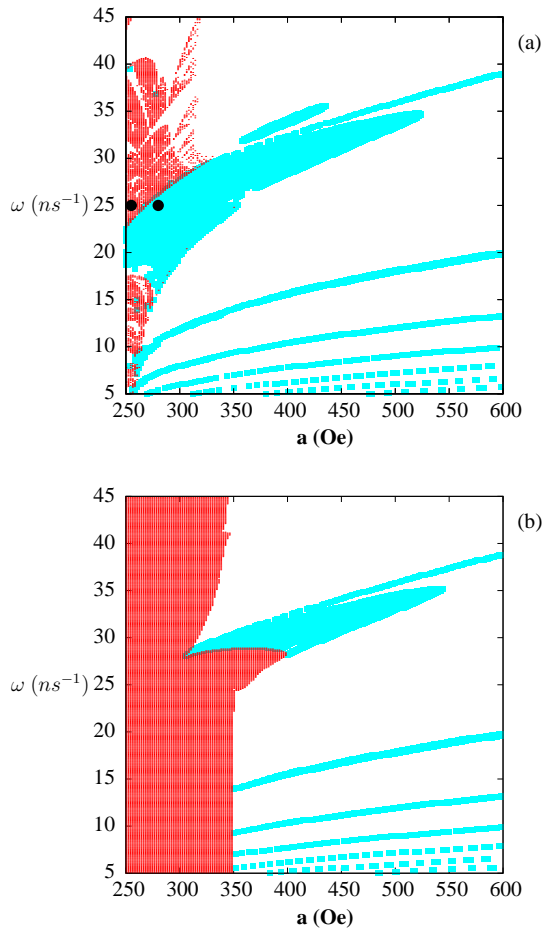}
\caption{Regions of {\it chaos}(red dots) and {\it periodicity}
(blue wings) in the 
parameter
space of DC current $`a'$ and frequency $`\omega'$, a) without anisotropy
and b) with anisotropy ($\kappa=45~Oe$) along the $z$ direction. 
The left over regions show {\it multiply periodic} behavior. 
All other parameters remain the same as in FIG 3.  The power spectrum at 
the two dark points in (a) $(255,25)$ and $(280,25)$ are shown in FIG 5 (a)
and (b), respectively. }
\end{figure}

\subsubsection{Case b: Non-zero anisotropy($\kappa=45~Oe$)}
Contrary to the observation in FIG 2a, in the
presence of a non-zero easy axis anisotropy,
chaos is noted at much lower values of the spin current and
DC applied field (FIG 2b). 
FIGs 3b shows
period doubling bifurcation scenario in the presence of anisotropy.
In FIG 3b the current magnitude is varied, keeping the magnetic field
strength and frequency of the AC component fixed. 
Similar period doubling
route to chaos is also noticed as a) the magnitude of the
steady magnetic field is varied (keeping current and frequency fixed), and 
when the frequency of the AC component of the applied 
magnetic field is changed (keeping the current and magnetic field strengths
constant). In either case, an easy
axis anisotropy of magnitude $\kappa=45~Oe$ is chosen along the $z$ direction,
which is also the polarization direction of the spin current. The figures
corresponding to these results are, however, not presented in here. 

\begin{figure}[b]
\centering\includegraphics[width=1\linewidth]{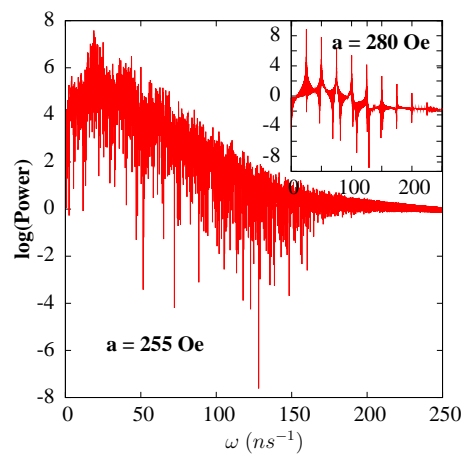}
\caption{The power spectrum distribution corresponding to periodic,
$a=280~Oe$ ({\it inset}), and  chaotic, $a=255~Oe$, scenarios in 
FIG 4a. The first peak in the inset is seen at $\omega=25~ns^{-1}$. The anisotropy is taken zero, and all other parameters are the same as
in FIG 4a.} 
\end{figure}

\begin{figure}[b]
\centering\includegraphics[width=1\linewidth]{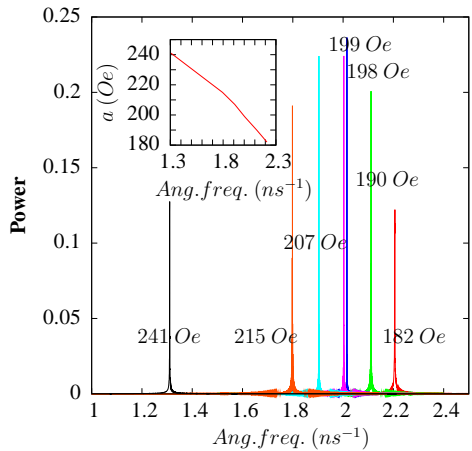}
\caption{The power spectrum distribution in the limit $\omega =0$, 
at certain values of $a$ (indicated on each spectrum) where 
periodic behavior is noted. Multiply periodic behavior is noticed for 
other values of $a$ in the range shown. 
The current magnitudes vary linearly and decrease with the frequency
of oscillation ({\it inset}). $h_{ac}=0$, while all other parameters are the 
same as in FIG. 3.  }
\end{figure}

\subsection{Periodic, Multiply periodic and Chaotic dynamics}

In the presence of an AC spin-current induced torque, it is known that as the 
frequency of the spin-current  is varied the system
exhibits three distinct phases wherein the dynamics is predominantly either - 
{\it Periodic, Multiply periodic} or {\it Chaotic}\cite{charles:2006}. Herein 
we show that an applied AC magnetic field of magnitude $10~Oe$, instead of
a AC current, results in the three dynamical phases 
as the frequency of the
applied field is varied for a constant value of the applied DC field.

FIG 4a shows regions of periodic (blue wings), and chaotic (red stem)
dynamics in the parameter space of the spin current magnitude and the
frequency of the applied magnetic field. Multiply periodic behavior is
seen in the unshaded regions. Here again the 
similarity with that of FIG 1 in \cite{charles:2006} may be noted. This
further illustrates the qualitative similarity of the spin current induced
torque with that of the applied field. 
 Our numerical results further show a number of wing like bands in 
the $\omega-a$ space where periodic behavior is noticed. This is clearly
absent with a periodic spin-current as seen in \cite{charles:2006}.
The power spectrum corresponding to two specific points, one chaotic and
the other periodic (indicated with 
dark dots in FIG 4a), are shown in FIG 5. Periodic behavior is noticed even
in the limit $\omega\to 0$ for certain values of $a$, while multiply 
periodic behavior is noticed for the other intermediate values. 
The power spectrum corresponding to these current values are shown in FIG. 6.
Such a behavior, induced by a spin-current of 
constant magnitude, has been noticed in \cite{ripp:2004}. Indeed, the 
current magnitudes, $a$, where periodic behavior is seen to vary linearly (with 
a negative slope) with the corresponding periods (see FIG. 6 inset), in further
agreement with \cite{ripp:2004}.

However, in the presence of the anisotropy field chaotic
regime is much more pronounced and wider, as seen in FIG 4b. For a much
lower value of the DC applied field ($h_{a3}=20~Oe$) there appear wide
bands in the $\omega-a$ space where periodic behavior is exclusively noted
at low frequencies (FIG 7). As the frequency is increased {\it multiple
periodicity islands} appear in between {\it periodic} bands, and for 
even higher frequencies the dynamics is largely one of multiply periodic 
type. No chaotic dynamics
is noted in the parameter range chosen. These thick periodicity bands
present better regions to choose in applications such as the microwave
resonator discussed earlier, while chaos synchronization studies can be
carried out in the chaos regimes. 
\begin{figure}[t]
\centering\includegraphics[width=1\linewidth]{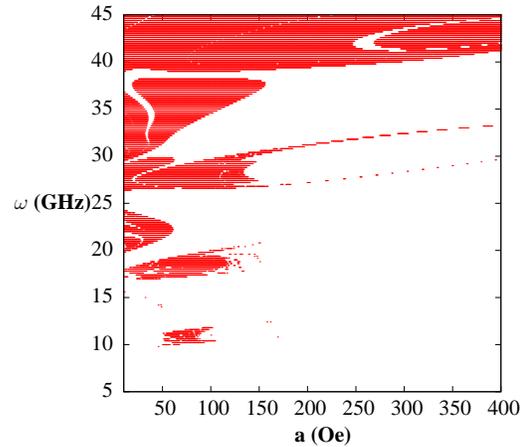}
\caption{Regions of {\it multiply periodic} dynamics 
for the system with the DC applied field fixed at $h_{a3}=20~Oe$, and
non-zero anisotropy.
All the other parameters remain
the same as in FIG 4b. Synchronization is noted in the unshaded regions, 
while chaotic 
dynamics is not
noticed in the parameter range shown in the figure. Islands of multiply
periodic behavior
appear between regions of periodic behavior for low frequencies. For 
higher frequencies, the dynamics is exclusively multiply periodic. }
\end{figure}

%The explicit time dynamics for the three scenarios for three
%sample values of $(a,\omega)$ is shown in FIG 6, by a plot of one 
%component of spin ($S_3$) versus time. This illustrates the nature the
%three dynamical scenarios. The dynamical behavior in the case
%of chaotic and periodic motion is evident from FIGs 5.a and 5.c. 
%A third behavior is a situation where the system shows a multiply periodic
%behavior, with small number of periods, but not leading to chaos as the
%parameter is changed.    

%\begin{figure}[h]
%\centering\includegraphics[width=1\linewidth]{fig7.eps}
%\caption{Explicit dynamics for the $S_2$ component of spin in time for the
%three different cases. a) Modification $a=300~Oe$ and $\omega=15~~ns^{-1}$,
%b) Chaos $a=250~Oe$, $\omega=30~~ns^{-1}$ and c) Synchronization $a=300~Oe$, 
%$\omega=25~~ns^{-1}$. }
%\end{figure}

\section{Discussion and conclusion}

Using the complex stereographic variable to represent the spin vector,
and rewriting the modified Landau-Lifshitz-Gilbert equation, we
have shown that the spin-current induced torque is qualitatively equivalent
to an applied field. Using this equivalence we have shown that an applied
AC magnetic field in the presence of constant spin current and DC applied 
magnetic field can lead to varied dynamical scenarios including {\it chaos}. 
We have explicitly demonstrated numerically the chaotic behavior for
a range of values of the parameters. The system also exhibits regular periodic
behavior for a different range of values. It is now realized that such 
nanoscale 
monodomain layers can find application as resonators, through periodic
oscillations induced by an alternating spin-current. The results presented
here provide an alternative method through oscillations induced by an applied 
magnetic field. 
It is further noticed that the range of the chaotic
regime strongly depends on the presence of a crystal anisotropy field. 
In the presence of an anisotropy field chaotic behavior is noticed for much
lower values of the DC field and spin-current, which are more suited
for chaos synchronization studies. However, there are regions
in the $\omega-a$ space where regular periodic motion is more robust and
presents a better alternative in applications. In a future study we
will present the possibility of chaos synchronization in spin-valve
structures.

\section*{ACKNOWLEDGMENTS}
S.M. wishes to thank DST, India, for funding through the FASTTRACK scheme. 
The work of M.L. 
forms part of a DST-IRHPA research project and is supported by a DST-Ramanna
fellowship.

\end{document}